\documentclass[12pt,a4paper]{article}

\usepackage{amsmath,amsfonts,amssymb,amsthm,mathtools}

\usepackage[english]{babel}
\usepackage[utf8]{inputenc}
\usepackage[T1]{fontenc}

\usepackage{latexsym}
\usepackage[pdftex]{graphicx}
\usepackage{epstopdf}

\usepackage{subfigure}
\usepackage{MnSymbol}

\usepackage{enumitem}

\usepackage{accents}

\usepackage{hyperref}
\usepackage[all]{hypcap}

\usepackage{doi}

\usepackage{fancyhdr}
\usepackage[hmarginratio=1:1]{geometry}

\usepackage{authblk}

\usepackage[normalem]{ulem}

\hypersetup{
	pdftitle={},
	pdfauthor={},
	colorlinks=true,
	linkcolor=black,
	citecolor=black,
	filecolor=black,
	urlcolor=blue,
	bookmarksopen,
	bookmarksopenlevel=1,
}

\newcommand{\cC}{\mathcal{C}}

\newcommand{\ee}{\mathrm{e}}

\newcommand{\dd}{\mathrm{d}}

\def\eps{\epsilon}

\theoremstyle{definition}

\theoremstyle{remark}

\begin{document}


\title{%
Stability of the classical catenoid and Darboux-P\"oschl-Teller potentials%
}%

\renewcommand\Authfont{\normalsize}
\author[1]{%
Jens Hoppe%
}%

\author[2]{%
Per Moosavi%
\thanks{\texttt{pmoosavi@phys.ethz.ch}}%
}%

\renewcommand\Affilfont{\footnotesize}
\affil[1]{%
Technische Universit{\"a}t Braunschweig, Brunswick, Germany%
}

\affil[2]{%
Institute for Theoretical Physics, ETH Zurich, Wolfgang-Pauli-Strasse 27, 8093 Z{\"u}rich, Switzerland%
\vspace{-10pt}
}%

\date{%
October 31, 2022%
\vspace{-30pt}
}%


\maketitle


\begin{abstract}
We revisit the stability (instability) of the outer (inner) catenoid connecting two concentric circular rings and give an explicit new construction of the unstable mode of the inner catenoid by studying the spectrum of an exactly solvable one-dimensional Schr\"odinger operator with an asymmetric Darboux-P\"oschl-Teller potential.
\end{abstract}


\section{Introduction}
\label{Sec:Introduction}


The catenoid is one of the most prominent and thoroughly discussed minimal surfaces, appearing already in the work of Euler.
Still, it is not always appreciated that its stability properties are closely linked to exactly solvable one-dimensional Schr{\"o}dinger operators.
Indeed, the stability operator of the catenoid, arising from the second variation of the area functional, is of Schr{\"o}dinger-type with a hyperbolic Darboux-P{\"o}schl-Teller (DPT) potential \cite{Darboux1882, PoschlTeller1933} and Dirichlet boundary conditions.
This is an example of an interesting connection between minimal surfaces and Schr{\"o}dinger operators, cf., e.g., \cite{GaillardMatveev2009, Veselov2011}.

\begin{figure}[!htbp]

\centering
\includegraphics[scale=1, trim=0mm 0mm 0mm 0mm, clip=true]{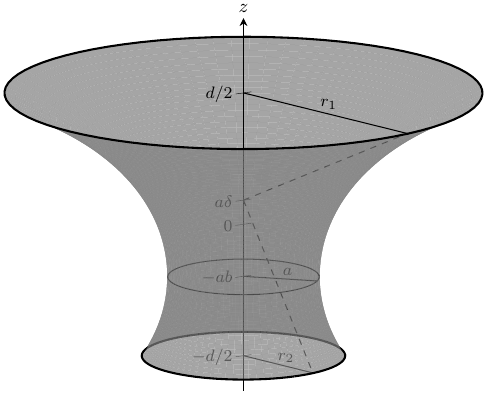}

\caption{%
Illustration of the critical catenoid connecting two given concentric circular rings.
See \eqref{BCs} and its preceding text as well as \eqref{delta_wc_b} for the quantities in the figure.
The dashed lines are tangent to the surface at the two boundary rings.%
}%

\label{Fig:Catenoid}

\end{figure}

Viewing the catenoid as the minimal surface connecting two given concentric circular rings, see Fig.~\ref{Fig:Catenoid}, one would like to determine when such catenoid solutions exist.
In general, for given boundary rings at a certain distance, there can be no, exactly one, or two solutions.
The latter are in the form of one outer and one inner catenoid, while the situation with one solution occurs at a particular, critical, distance.
In the case of rings with the same radii, it is a textbook exercise to obtain the critical catenoid beyond which there are no solutions.
This special case is known to yield a symmetric hyperbolic DPT potential when looking at the second variation.
Using the factorization method, well known in the theory of integrable systems, one can derive an explicit formula for the ground state for the critical catenoid and in this way study its stability (as explained, e.g., in \cite{Hoppe2019}).
See also, e.g., \cite{BerardSaEarp2010, TamZhou2009} and references therein for results in other geometries and in higher dimensions.

The general case of different radii, although discussed already one to two centuries ago \cite{Goldschmidt1831, Lindelof1869}, is somewhat more complicated and leads to an asymmetric potential.
Fortunately, even this case is amenable to exact analytical solutions:
The spectrum of the corresponding Schr{\"o}dinger-type stability operator can again be studied using the factorization method and an explicit expression for a zero-energy eigenstate of the stability operator (at criticality) obtained.
By showing that this expression is positive, thus the ground state, we prove the stability (instability) of the outer (inner) catenoid for general boundary rings.
Remarkably, the factorization method also allows us to explicitly construct the unstable mode of the inner catenoid, for which we are not aware of any geometric construction.

This paper is organized as follows.
In Sect.~\ref{Sec:Existence} we study the existence of catenoid solutions for given boundary rings and construct ``phase diagrams'' with a critical curve above (below) which such solutions do (not) exist.
In Sect.~\ref{Sec:Stability} we prove the stability (instability) of the outer (inner) catenoid above the critical curve.
In Sect.~\ref{Sec:Goldschmidt_solution} we discuss the Goldschmidt solution (which consists of two disks covering the circular rings connected by a line along the $z$-axis) and show that the outer catenoid becomes meta-stable before it becomes unstable.
In Sect.~\ref{Sec:Critical_distance} we derive a formula for the critical distance between the two boundary rings for given radii.
The unstable mode of the inner catenoid is constructed in Sect.~\ref{Sec:Unstable_mode_of_the_inner_catenoid}.


\section{Existence of catenoid solutions}
\label{Sec:Existence}


Consider the area $A$ of the axially symmetric surface connecting two concentric circular rings of radii $r_{1}$ and $r_{2}$, respectively, in parallel planes separated by a distance $d$.
Varying the shape of the surface, one finds that stationary points of $A$ arise if and only if the radius $r$ as a function of the height $z$ is of the form
\begin{equation}
\label{r_z}
r(z) = a \cosh(z/a + b),
\end{equation}
where the constants $a$ and $b$ as functions of the data $r_{1}$, $r_{2}$, and $d$ are determined by the boundary conditions
\begin{equation}
\label{BCs}
\begin{aligned}
r_{1}
& = a \cosh(d/2a + b), \\
r_{2}
& = a \cosh(d/2a - b).
\end{aligned}
\end{equation}
(Without loss of generality, $z = 0$ is assumed to be the plane that has equal distance to the two rings.)
Rewriting \eqref{BCs} as
\begin{subequations}
\label{rho_xi_def}
\begin{align}
\rho
& := \frac{r_{1} + r_{2}}{d}
	= \frac{\cosh(w)}{w} \cosh(b),
	\label{rho_xi_def_1} \\
\xi
& := \frac{r_{1} - r_{2}}{d}
	= \frac{\sinh(w)}{w} \sinh(b),
	\label{rho_xi_def_2}
\end{align}
\end{subequations}
where $w := d/2a$, one finds
\begin{equation}
\label{h_function}
\rho w = \cosh(w) \sqrt{ 1 + \frac{w^2 \xi^2}{\sinh^2(w)} } =: h(w).
\end{equation}
These equations determine $w$ in terms of $(r_{1}/d, r_{2}/d)$ or $(\xi, \rho)$, respectively.%
\footnote{%
Recall that minimal surfaces are scale invariant.
Thus, the catenoid only depends on dimensionless ratios of $r_{1}$, $r_{2}$, and $d$, which is equivalent to using $\rho$ and $\xi$.%
}
As $h(w)$ is monotonic in $w > 0$, starting at $h(0) = \sqrt{1 + \xi^2}$, the equation in \eqref{h_function} will have no solution for small $\rho$ ($< \rho_{c}$), two solutions,
$w_{1}$ and $w_{2}$, for large $\rho$ ($> \rho_{c}$), and exactly one solution, $w_{c}$, if the linear function $\rho w$ just touches the curve $h(w)$, which happens at some critical value, $\rho = \rho_{c}$.
Above criticality, $w_{1} < w_{c} < w_{2}$ due to monotonicity of $h(w)$, where $w_{2}$ corresponds to the inner catenoid and $w_{1}$ to the outer catenoid since the associated smallest distances $a_{1,2} = d/2w_{1,2}$ to the $z$-axis satisfy $a_{2} < a_{1}$.
In addition, inserting $w_{1,2}$ into \eqref{BCs} or \eqref{rho_xi_def} [for given $(r_{1}/d, r_{2}/d)$ or $(\xi, \rho)$, respectively] yields the corresponding values $b_{1,2}$.

For the critical case, differentiating $h(w)$ with respect to $w$ yields one more condition, which together with \eqref{h_function} implies
\begin{subequations}
\label{rho_xi_w}
\begin{align}
\rho^2 w_{c}^3
& = \cosh^4 (w_{c}) [w_{c} - \tanh(w_{c})],
	\label{rho_xi_w_1} \\
\xi^2 w_{c}^3
& = \sinh^4 (w_{c}) [w_{c} - \coth(w_{c})].
	\label{rho_xi_w_2}
\end{align}
\end{subequations}
Reintroducing the constant $b$, this yields
\begin{equation}
\label{wc_b}
2 w_{c} = \coth(w_{c} - b) + \coth(w_{c} + b),
\end{equation}
determining $w_{c}$ (only) in terms of the dimensionless real constant $b$,%
\footnote{%
Note that $w_{c}$ goes as $b + 1/(2b-1) + \ldots$ for large $b > 0$.%
}
or put differently, $\cosh(2b) = \cosh(2w_{c}) - {\sinh(2w_{c})}/{w_{c}}$, which implies that for each possible $w_{c}$ there is a unique value of $|b|$.
The special case $b = 0$ corresponds to $r_{1} = r_{2}$, i.e., $\xi = 0$, cf.\ \eqref{rho_xi_def_2} and \eqref{rho_xi_w_2},
implying $\rho_{c} = \rho_{0}$ and $w_{c} = w_{0}$ with $\rho_{0} = \sinh(w_{0})$ and $w_{0} \approx 6/5$ solving $w_{0} = \coth(w_{0})$,%
\footnote{%
\label{Footnote:rho_0_w_0}%
For $\xi = 0$, \eqref{h_function} can be written as
$\rho = w^{-1} \cosh(w) =: \rho(w)$.
This equation has exactly one solution $w = w_{0}$ when $\rho = \rho_{0}$ for
$\rho_{0} = \rho_{0}(w_{0}) : = 1/\sqrt{w_{0}^2 - 1} \approx 3/2$
and $w_{0} \approx 6/5$ given by the solution to
$0 = (\dd \rho(w) / \dd w) \big|_{w = w_{0}}
= w^{-2} \sinh(w) [w - \coth(w)] \big|_{w = w_{0}}$.%
}
while $w_{c} \geq w_{0}$ in general.
Note that when inserting \eqref{rho_xi_def} into \eqref{rho_xi_w}, the difference of the two equations becomes a tautology while their sum gives \eqref{wc_b}.
For later reference, we also note that \eqref{rho_xi_w} can be viewed as a curve in the $(\xi, \rho)$-half-plane parametrized by $w_{c} \in [ w_{0}, \infty)$.

\begin{figure}[!htbp]

\centering

\subfigure[``Phase diagram'' in $(\xi, \rho)$-space.]{
\centering
\includegraphics[scale=0.7]{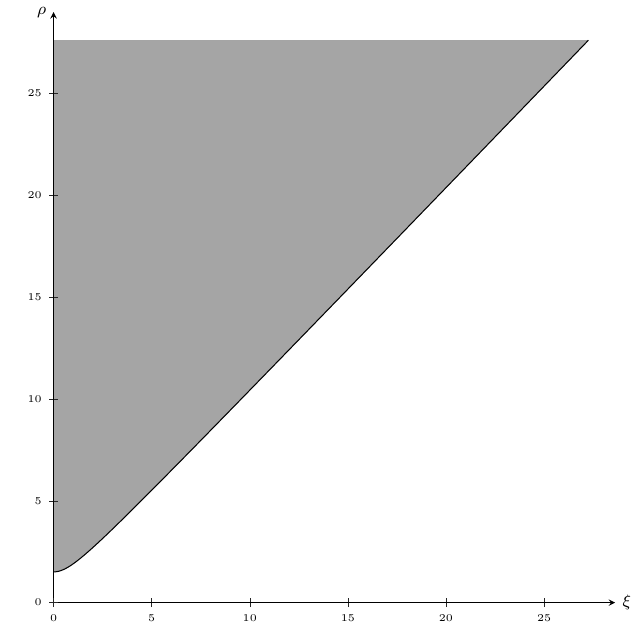}
\label{Fig:Phase_diagram:xi_rho}
}
\subfigure[``Phase diagram'' in $(b, w)$-space.]{
\centering
\includegraphics[scale=0.7]{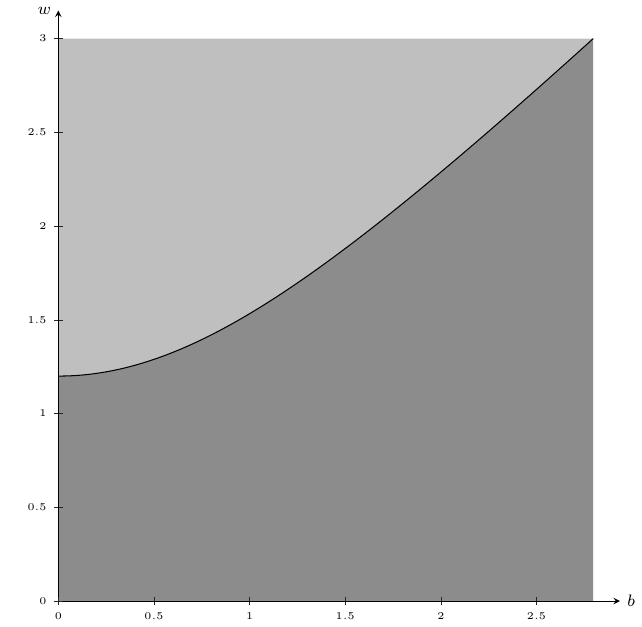}
\label{Fig:Phase_diagram:b_w}
}

\caption{%
``Phase diagrams'' in \subref{Fig:Phase_diagram:xi_rho} $(\xi, \rho)$-space for $\xi, \rho \geq 0$ and \subref{Fig:Phase_diagram:b_w} $(b, w)$-space for $b, w \geq 0$.
(The diagrams are symmetric under $\xi \to -\xi$ and $b \to -b$, respectively.)
In \subref{Fig:Phase_diagram:xi_rho}, the grey region corresponds to when the equation system in \eqref{BCs} can be solved and the white region to when it cannot.
In \subref{Fig:Phase_diagram:b_w}, the dark (light) grey region corresponds to outer- (inner-) catenoid solutions.
The black critical curves are given by \eqref{rho_xi_w} and \eqref{wc_b}, respectively.%
}%
	
\label{Fig:Phase_diagram}
	
\end{figure}

Another way to derive (and view) \eqref{wc_b} is to compute the Jacobian of the map
$\mathbb{R}^{+} \times \mathbb{R} \ni (b, w)
\mapsto (\xi, \rho) \in \mathbb{R}^{+} \times \mathbb{R}$
given by \eqref{rho_xi_def}:%
\footnote{%
Defining
$\rho_{+} := 2r_{1}/d = w^{-1} \cosh(w + b)$
and
$\rho_{-} := 2r_{2}/d = w^{-1} \cosh(w - b)$,
the curve \eqref{wc_b} is actually more simply given by 
\begin{equation*}
\left| \frac{\partial(\rho_{+}, \rho_{-})}{\partial(b, w)} \right|
= \frac{1}{w^3}
	\bigl[
		\sinh(2w) - 2w\sinh(w + b) \sinh(w - b) 
	\bigr]
= 0.
\end{equation*}
}
\begin{equation}
\left| \frac{\partial(\xi, \rho)}{\partial(b, w)} \right|
= \frac{1}{w^2}
		\left[
			\cosh^2(b) \sinh^2(w) - \sinh^2(b) \cosh^2(w)
		\right]
		- \frac{\sinh(w) \cosh(w)}{w^3}.
\end{equation}
Its vanishing gives the same critical curve in the $(b, w)$-half-plane as \eqref{wc_b}, which is easily seen by using $\coth(w \pm b) = [1 \pm \coth(w) \coth(b)]/[\coth(w) \pm \coth(b)]$.
This critical curve $\cC$ is mapped by \eqref{rho_xi_def} onto a critical curve $\tilde{\cC}$ in the $(\xi, \rho)$-half-plane, which has the property that on one side of $\tilde{\cC}$ there is no solution to \eqref{rho_xi_def}, two solutions on the other side, and exactly one solution for each given $(\rho_{c}, \xi_{c}) \in \tilde{\cC}$, see Fig.~\ref{Fig:Phase_diagram:xi_rho}.
The corresponding curve $\cC$ in the $(b, w)$-half-plan instead separates outer- and inner-catenoid solutions, see Fig.~\ref{Fig:Phase_diagram:b_w}.%
\footnote{%
As a side remark, we mention that similar ideas as those presented here, specifically the derivation of \eqref{rho_xi_w}, were recently shown to be useful for studying Floquet drives in inhomogeneous conformal field theory \cite{LapierreMoosavi2021}.%
}

%


\section{Stability analysis}
\label{Sec:Stability}


Just as the vanishing of the first variation of the area functional gave \eqref{r_z}, it is straightforward to see that, upon perturbing \eqref{r_z} by a function $\eps(z) = a \tilde{\eps}(v)$, $v := z/a$, satisfying $\tilde{\eps}(\pm w) = 0$,%
\footnote{%
Because of the boundary conditions, we require that $\eps(z)$ vanishes at $z = \pm d/2$.%
}
the second variation is
\begin{equation}
\delta^2 A
= \pi a^2 \int_{-w}^{w}
	\frac{\tilde{\eps}(v)}{\cosh(v + b)}
	J_{b}
	\frac{\tilde{\eps}(v)}{\cosh(v + b)}
	\, \dd v,
\end{equation}
with the operator
\begin{equation}
\label{J_operator}
J_{b} := - \partial_{v}^2 - \frac{2}{\cosh^2(v + b)}.
\end{equation}
This operator acts on square-integrable functions on $I := (-w, w)$ that vanish on the boundary of the interval.

We want to prove that on the critical curve $\cC$, i.e., choosing $I = I_{c} = (-w_{c}, w_{c})$ with $w_{c} = w_{c}(b)$ given by \eqref{wc_b}, $J_{b}$ has an eigenvalue zero with strictly positive eigenfunction $\Psi(v)$ in $I_{c}$ that vanishes on the boundary.
Since $\Psi(v)$ then represents the ground state of $J_{b}$ by the Sturm-Liouville theorem, it follows that the lowest eigenvalue of the critical operator $J_{b}$ acting on $I_{c}$ is zero.
Hence, as the spectrum of $J_{b}$ goes up when reducing the interval and down when enlarging it, this proves that the outer catenoid ($w = w_{1} < w_{c}$) is stable while the inner catenoid ($w = w_{2} > w_{c}$) is unstable.

The function $\Psi(v)$ can be found explicitly using the factorization method as follows.
Noting that
\begin{equation}
J_{b}
= [- \partial_{v} + \tanh(v + b)][\partial_{v} + \tanh(v + b)] - 1
= L^{\dagger} L - 1
\end{equation}
with $L = \partial_{v} + \tanh(v + b)$, we can form
\begin{equation}
\tilde{J}_{b} := L L^{\dagger} - 1 = - \partial_{v}^2 =: \tilde{J}.
\end{equation}
[In the language of supersymmetric quantum mechanics, $J_{b} + 1$ and $\tilde{J} + 1$ correspond to supersymmetric partner Hamiltonians, and the supersymmetric partner potential of $-2/\cosh^2(v+b)$ is constant, i.e., equivalent to the free case.]
Since $\tilde{J}$ acting on $-v + \delta$ is zero for any real constant $\delta$, it follows that $J_{b} L^{\dagger}(-v + \delta) = 0$, and thus
\begin{equation}
\label{Psi_v_GS}
\Psi(v) := 1 + (-v + \delta) \tanh(v + b)
\end{equation}
is annihilated by $J_{b}$.%
\footnote{%
\label{Footnote:Psi_geometric_interpretation}%
In differential geometry (cf.\ the footnote on p.\ 4 of \cite{Hoppe2019}) the zero-energy eigenfunction of the critical catenoid is known as the projection, onto the surface normal, of the position vector -- taking as the origin the point $(0, a\delta)$ on the symmetry axis where the lines tangent to the surface at the two boundary rings meet [cf.\ \eqref{tangent_construction} and Fig.~\ref{Fig:Catenoid}].%
}
Requiring that $\Psi(\pm w) = 0$, we obtain the two conditions
\begin{equation}
\label{delta_w_b}
\delta = w - \coth(w + b),
\qquad
\delta = - w + \coth(w - b),
\end{equation}
which at criticality are consistent with \eqref{wc_b}, as well as allows one to write $\delta$ in the more symmetric form
\begin{equation}
\label{delta_wc_b}
\delta
= \frac{1}{2} \left[ \coth(w_{c} - b) - \coth(w_{c} + b) \right],
\end{equation}
or equivalently, $\delta = \sinh(2b)/[\cosh(2w_{c}) - \cosh(2b)]$.

It remains to prove that $\Psi(v)$ in \eqref{Psi_v_GS} is strictly positive in $I_{c}$.
This can be done by contradiction as follows:
Assume $\Psi(v) > 0$ is not true everywhere in $I_{c}$.
Given that $\Psi(\pm w_{c}) = 0$, then $\Psi(v) = 0$ also somewhere inside $I_{c}$.
Since $\Psi(v)$ is differentiable, it follows that it must have more than one point with zero derivative.
Solving
$0 = \dd \Psi(v)/\dd v
= - \tanh(v + b) + (-v + \delta)/\cosh^2(v + b)
= [- \sinh(2[v + b]) + 2(-v + \delta)]/2\cosh^2(v + b)$
on the other hand implies
\begin{equation}
\label{deriv_zero}
\sinh(2[v + b]) + 2(v - \delta) = 0.
\end{equation}
The left-hand side is monotonically increasing, meaning that it can only have one zero, which corresponds to a maximum of $\Psi(v)$ since $\Psi'(- w_{c}) > 0$ and $\Psi'(w_{c}) < 0$.
This contradicts the assumption, which means that
$\Psi(v) > 0$ everywhere in $I_{c}$ for all $b$.

As a last remark, \eqref{deriv_zero} implies that, unless $b = 0$, the maximum of $\Psi(v)$ is not located at the minimum $v = -b$ of the potential term in \eqref{J_operator} nor at the center of $I_{c}$ but rather at some point between $-b$ and $0$.
This is a direct consequence of the asymmetric potential since it leads to a competition between minimizing the kinetic- and potential-energy terms in \eqref{J_operator}.


\section{Goldschmidt solution and meta-stability}
\label{Sec:Goldschmidt_solution}


The surface area of the catenoid can be computed from $r(z)$ in \eqref{r_z} as
\begin{equation}
\label{A_C}
A_{\mathrm{C}}
= 2\pi \int_{-d/2}^{d/2} \dd z\, r(z) \sqrt{1 + r'(z)^2}
= \pi d^2
	\Biggl[
		\frac{a}{d}
		+ \biggl( \frac{a}{d} \biggr)^2 \sinh(d/a) \cosh(2b)
	\Biggr].
\end{equation}
Using $\rho$ and $\xi$ introduced in \eqref{rho_xi_def},
or alternatively using only $w$ and $b$, this can be written
\begin{equation}
\label{A_C_ver2}
A_{\mathrm{C}}
= \frac{\pi d^2}{2}
	\biggl(
		\frac{1}{w} + \rho^2 \tanh(w) + \xi^2 \coth(w)
	\biggr)
= \frac{\pi d^2}{4 w^2}
	\bigl( 2w + \sinh(2w) \cosh(2b) \bigr).
\end{equation}
This should be compared with the area of the solution that is not a catenoid but instead consists of disks whose boundaries are the two circular rings, which together with the line along the $z$-axis connecting the disks is the Goldschmidt solution \cite{Goldschmidt1831}:
Such a solution appears as the surface of revolution of a discontinuous function of $z$ that equals $r_{1}$ and $r_{2}$ at $z = \pm d/2$, respectively, and zero elsewhere.
The total area of these disks is
\begin{equation}
\label{A_G}
A_{\mathrm{G}}
= \pi \bigl( r_{1}^2 + r_{2}^2 \bigr)
= \frac{\pi d^2}{2}
	\bigl( \rho^2 + \xi^2 \bigr)
= \frac{\pi d^2}{4 w^2}
	\bigl( 1 + \cosh(2w) \cosh(2b) \bigr),
\end{equation}
and its difference with the catenoid can therefore be written
\begin{equation}
\label{A_C_m_A_G}
A_{\mathrm{C}} - A_{\mathrm{G}}
= \frac{\pi d^2}{2}
	\biggl(
		\frac{1}{w}
		+ \rho^2 \bigl[ \tanh(w) - 1 \bigr]
		+ \xi^2 \bigl[ \coth(w) - 1 \bigr]
	\biggr)
= \frac{\pi d^2}{4 w^2}
	\bigl( 2w - 1 - \cosh(2b) \ee^{-2w} \bigr).
\end{equation}
For the critical curve given by \eqref{rho_xi_w} parametrized by $w_{c} \in [ w_{0}, \infty)$, or alternatively using \eqref{wc_b}, one obtains
\begin{equation}
A_{\mathrm{C}} - A_{\mathrm{G}}
= \frac{\pi d^2}{2} f(w_{c}),
\qquad
f(w_{c})
= \frac{1}{w_{c}} - \frac{3}{4 w_{c}^2} + \frac{1}{4w_{c}^3}
	- \frac{\ee^{-4w_{c}}}{4 w_{c}^2} - \frac{\ee^{-4w_{c}}}{4 w_{c}^3},
\end{equation}
which implies
$A_{\mathrm{C}} > A_{\mathrm{G}}$
since $f(w_{c}) > 0$ for $w_{c} \geq w_{0}$.
(One can convince oneself of the latter by plotting $f(w_{c})$ up to some large but finite $w_{c}$ and by noting that the explicit formula approaches $0$ from above as $w_{c} \to \infty$.)
Thus, the area of the Goldschmidt solution becomes smaller than that of the catenoid before the equation system in \eqref{BCs} ceases to have a solution.

Since we proved in Sect.~\ref{Sec:Stability} that the outer catenoid is stable above the critical curve in Fig.~\ref{Fig:Phase_diagram:xi_rho}, it follows from the above that it actually becomes meta-stable before it becomes unstable.
More precisely, increasing the distance between the given boundary rings, the Goldschmidt solution, as is well known, becomes the global area minimum before the catenoid becomes critical.

Lastly, by using \eqref{rho_xi_def} to rewrite \eqref{A_C_ver2} as
\begin{equation}
\label{A_C_ver3}
A_{\mathrm{C}}
= \frac{\pi d^2}{2w}
	\biggl(
		1 + \rho \sinh(w) \cosh(b) + \xi \cosh(w) \sinh(b)
	\biggr),
\end{equation}
one can show that $A_{\mathrm{C}, 2} > A_{\mathrm{C}, 1}$, where $A_{\mathrm{C}, 1}$ ($A_{\mathrm{C}, 2}$) is the surface area of the outer (inner) catenoid corresponding to $w_{1}$ ($w_{2}$).
Indeed, for given $(r_{1}, r_{2}, d)$ [or $(\xi, \rho, d)$], it follows from \eqref{A_C_ver3}, \eqref{rho_xi_def}, and straightforward manipulations that
\begin{equation}
\frac{2 w_{1} w_{2}}{\pi d^2} (A_{\mathrm{C}, 2} - A_{\mathrm{C}, 1})
= \sinh(w_{2} - w_{1}) \cosh(b_{2} - b_{1}) - (w_{2} - w_{1})
> 0
\end{equation}
since $\sinh(w_{2} - w_{1}) > w_{2} - w_{1} > 0$ and $\cosh(b_{2} - b_{1}) \geq 1$.


\section{Critical distance}
\label{Sec:Critical_distance}


Given $r_{1}$ and $r_{2}$, the critical distance $d_{c}(r_{1}, r_{2})$ between the boundary rings [beyond which the equation system in \eqref{BCs} has no solution] is determined by the critical curve in \eqref{rho_xi_w} in the $(\xi, \rho)$-half-plane parametrized by $w_{c}$.
Using this parametrization, it is possible to implicitly compute $w_{c}$ as a function of $r_{1}$ and $r_{2}$ as follows:
\begin{equation}
\label{g_function}
\frac{r_{1} - r_{2}}{r_{1} + r_{2}}
= \xi/\rho
= \tanh^{2}(w_{c}) \sqrt{ \frac{w_{c} - \coth(w_{c})}{w_{c} - \tanh(w_{c})} }
=: g(w_{c}).
\end{equation}
In the special case $r_{1} = r_{2}$, one has to solve $w_{c} = \coth(w_{c})$, whose solution we denoted by $w_{0}$, and \eqref{g_function} generalizes this.
It is straightforward to show that $g'(w_{c}) > 0$ for $w_{c} \geq w_{0}$, i.e., $g(w_{c})$ is monotonically increasing, and by inverting this function we can express the critical distance as
\begin{equation}
\label{d_c}
d_{c}(r_{1}, r_{2})
= \frac{r_{1} + r_{2}}{\cosh^2(w_{c})} \frac{w_{c}^{3/2}}{\sqrt{w_{c} -  \tanh(w_{c})}},
\qquad
w_{c} = g^{-1} \left( \frac{r_{1} - r_{2}}{r_{1} + r_{2}} \right).
\end{equation}

We note that $r_{1} = r_{2}$ yields the largest critical distance for a given value of $\max(r_{1}, r_{2})$.
Moreover, the critical distance depends linearly on $r_{1}$ for a given ratio $r_{1}/r_{2}$ since the corresponding parameter $w_{c}$ in \eqref{d_c} is then constant.

As is perhaps not generally known, the critical distance was studied already in \cite{Lindelof1869}:
Lindel{\"o}f gave an argument that the critical catenoid corresponds to the situation where the tangents to the surface at $z = \pm d/2$ meet on the symmetry axis.
This construction can be shown to be equivalent to \eqref{wc_b}.
Indeed, it follows from \eqref{r_z} that the slopes of these tangents are $r'(\pm d/2) = \pm \sinh(w \pm b)$, and thus that the sum of their projection onto the symmetry axis is
\begin{equation}
\label{tangent_construction}
\frac{r_{1}}{\sinh(w + b)} + \frac{r_{2}}{\sinh(w - b)}
= a [ \coth(w + b) + \coth(w - b)],
\end{equation}
where we used \eqref{BCs}.
This equals exactly $d = 2aw_{c}$ when \eqref{wc_b} is satisfied, cf.\ Fig.~\ref{Fig:Catenoid}.
See also \cite{BerardSaEarp2010} for the recovery of Lindel{\"o}f's result using Jacobi fields.


\section{Unstable mode of the inner catenoid}
\label{Sec:Unstable_mode_of_the_inner_catenoid}


Consider now the inner catenoid, i.e., $w = w_{2} > w_{c}$ and $b = b_{2}$ solving \eqref{rho_xi_def} for given $(r_{1}/d, r_{2}/d)$ [or $(\xi, \rho)$].
For such a given pair $(b, w)$,
\begin{equation}
\label{J_b_k2}
J_{b} \Psi_{k}(v) = - k^2 \Psi_{k}(v)
\end{equation}
is satisfied by
\begin{align}
\Psi_{k}(v)
& = [- \partial_{v} + \tanh(v + b)]
		\left( - \frac{\sinh(kv)}{k} + \beta(k) \cosh(kv) \right) \nonumber \\
& = \cosh(kv) - k\beta(k) \sinh(kv)
		+ \tanh(v + b) \left( - \frac{\sinh(kv)}{k} + \beta(k) \cosh(kv) \right),
		\label{Psi_k}
\end{align}
where $v \in (-w, w)$ and $J_{b}$ is as in \eqref{J_operator} with $\beta(k)$ (and $k$) determined by the boundary conditions
\begin{equation}
\label{Psi_k_BC}
\Psi_{k}(\pm w) = 0.
\end{equation}

In the regime where $\eps = w - w_{c} > 0$ is small, we write
\begin{equation}
\beta(k) = \delta + \gamma k^2 + O(k^4)
\end{equation}
with $\delta$, $\gamma$, and $k^2$ determined by \eqref{Psi_k_BC}.
We consider this regime in what follows and omit terms of higher order than $k^2$: This is justified below by showing that $\gamma = O(1)$ and $k^2 = O(\eps)$, while $\delta$ is independent of $\epsilon$.
Note that, for given $(r_{1}/d, r_{2}/d)$ [or $(\xi, \rho)$], one should view $b = b(\epsilon)$ as a function of $\epsilon$ according to \eqref{rho_xi_def}, with $b(\eps) = 0$ identically in the special case $r_{1} = r_{2}$ [or $\xi = 0$].

The dependence of $\Psi_{k}(v)$ on $k$ is smooth and such that the results for the critical catenoid are recovered as $k \to 0$.
It follows that $\Psi_{k}(v)$ represents the ground state of $J_{b}$ for the inner catenoid and that it is unstable since the corresponding eigenvalue is negative.%
\footnote{%
In general, there could be several values of $k$ such that \eqref{J_b_k2} and \eqref{Psi_k_BC} are satisfied, but for $\eps$ sufficiently small there will be only one.%
}

As in Sect.~\ref{Sec:Stability}, $\delta$ is given by \eqref{delta_wc_b}, which corresponds to $k = 0$, and thus to criticality.
Away from criticality, \eqref{Psi_k_BC} implies
\begin{subequations}
\begin{align}
c_{+} - w + \delta
+ \frac{k^2}{2}
	\left[
		w (w - 2\delta) c_{+} 
		- \frac{w^3}{3} + \delta w^2 + 2\gamma
	\right] + O(k^4) & = 0, \\
c_{-} - w - \delta
+ \frac{k^2}{2}
	\left[
		w (w + 2\delta) c_{-} 
		- \frac{w^3}{3} - \delta w^2 - 2\gamma
	\right] + O(k^4) & = 0,
\end{align}
\end{subequations}
where $c_{\pm} := \coth(w \pm b)$, and thus
\begin{equation}
\label{k2_two_formulas}
k^2
= \frac{6(c_{+} - w + \delta)}{w^3 - 3\delta w^2 - 6\gamma - 3w (w - 2\delta) c_{+}},
\qquad
k^2
=
\frac{6(c_{-} - w - \delta)}{w^3 + 3\delta w^2 + 6\gamma - 3w (w + 2\delta) c_{-}},
\end{equation}
up to higher-order corrections.
For these to be consistent, it follows that
\begin{subequations}
\label{gamma_and_gamma0}
\begin{align}
\gamma
& = \frac{w}{3}
		\frac{
			2 \delta (w^2 + 3c_{+} c_{-})
			+ (w^2 - 3\delta w - 3\delta^2) c_{+}
			- (w^2 + 3\delta w - 3\delta^2) c_{-}
		}{
			c_{+} + c_{-} - 2w
		}
	= \gamma_{0} + O(\eps),
	\label{gamma} \\
\gamma_{0}
& = \frac{w_{c}}{3}
		\frac{
			\delta w_{c}(w_{c}^2 - 3\delta^2)
			+ (w_{c}^4 - 2\delta^2 w_{c}^2 + 3\delta^4 - w_{c}^2 - 3 \delta^2) b'(0) 
		}{
			w_{c}^2 + \delta^2 - 2\delta w_{c} b'(0)
		}
	= \frac{\delta (1 - \delta^2)}{3},
	\label{gamma0}
\end{align}
\end{subequations}
where the second step in \eqref{gamma} follows from expanding in $\eps$ as well as using \eqref{wc_b} and \eqref{delta_wc_b},%
\footnote{%
Note that
$c_{\pm} = w_{c} \mp \delta + \eps [1 - (w_{c} \mp \delta)^2] [1 \pm b'(0)] + O(\eps^2)$.%
}
while the second step in \eqref{gamma0} follows from $b'(0) = -\delta / w_{c}$, which in turn is obtained by expanding in $\eps$ either \eqref{BCs}, written as $2r_{1}/d = w^{-1} \cosh(w + b)$ and $2r_{2}/d = w^{-1} \cosh(w - b)$,%
\footnote{%
Expanding these expressions for $2r_{1}/d$ and $2r_{2}/d$ in $\eps$ is also a simple way to derive \eqref{wc_b}.%
}
or the ratio of \eqref{rho_xi_def_2} and \eqref{rho_xi_def_1}.
Similarly, by expanding either of the two formulas in \eqref{k2_two_formulas} in $\eps$ and using \eqref{gamma_and_gamma0}, one obtains
\begin{equation}
\label{k2}
k^2
= \frac{3 \eps}{w_{c}} + O(\eps^2).
\end{equation}
Clearly, $k^2$ is proportional to $\eps$ and $\gamma = O(1)$, which justifies our series expansion in $k^2$.

It is remarkable that \eqref{k2} generalizes in such a straightforward way the corresponding result for the special case $r_{1} = r_{2}$ \cite{Hoppe2019}.
In that case, $b = 0$ and $\beta(k) = 0$, which inserted into \eqref{k2_two_formulas} and \eqref{Psi_k} yields
\begin{equation}
\label{k2_special_case}
k^2
= \frac{6}{w^2} \frac{\coth(w) - w}{w - 3\coth(w)}
= \frac{3 \eps}{w_{0}} + O(\eps^2),
\end{equation}
where the last step follows from $w = w_{2} = w_{0} + \eps$, and
\begin{equation}
\label{Psi_k_special_case}
\Psi_{k}(v)
= \cosh(kv) - \tanh(v) \frac{\sinh(kv)}{k}.
\end{equation}
Since $w_{c} = w_{0}$ when $r_{1} = r_{2}$, \eqref{k2_special_case} is the same as the result obtained directly from \eqref{k2}.

Generally, it would be interesting to see a geometric construction of the unstable mode.


\paragraph{Acknowledgments:}
P.M.\ gratefully acknowledges financial support from the Wenner-Gren Foundations (Grant No.\ WGF2019-0061).


\let\oldbibliography\thebibliography
\renewcommand\thebibliography[1]{
  \oldbibliography{#1}
  \setlength{\parskip}{0pt}
  \setlength{\itemsep}{0pt + 1.0ex}
}



\end{document}